\documentclass[ArXiv,AXISWP,accept,moreauthors,pdftex,10pt,letterpaper]{Definitions/axis} 
\usepackage{amssymb}

\firstpage{1} 
\pubvolume{xx}
\issuenum{1}
\articlenumber{5}
\pubyear{2023}
\copyrightyear{2023}

\newcommand{\fcgs}{erg s$^{-1}$ cm$^{-2}$}
\newcommand{\lx}{erg s$^{-1}$~}
\newcommand{\ms}{M$_\odot$}

\newcommand{\arcsec}{$^{\prime\prime}$~}

\Title{Surveying the onset and evolution of supermassive black holes at high-z with AXIS}

\Author{Nico Cappelluti $^{1}$, Adi Foord  $^{2}$ Stefano Marchesi $^{3,4,5}$, Fabio Pacucci $^{6,7}$, Angelo Ricarte $^{7,6}$,  Melanie Habouzit $^{10,11}$, Fabio Vito$^{5}$, Meredith Powell$^{2}$, Michael Koss$^{8}$, Richard Mushotzky$^{9}$, and the AXIS AGN-SWG}

\AuthorNames{Firstname Lastname, Firstname Lastname and Firstname Lastname}

\address{%
$^{1}$ \quad Department of Physics, University of Miami, Coral Gables, FL 33124, USA; ncappelluti@miami.edu\\
$^{2}$ \quad Department of Physics, University of Maryland Baltimore County, 1000 Hilltop Cir, Baltimore, MD 21250, USA \\
$^{3}$ \quad Department of Physics and Astronomy, Clemson University,  Kinard Lab of Physics, Clemson, SC 29634, USA \\
$^{4}$ \quad Dipartimento di Fisica e Astronomia (DIFA), Università di Bologna, via Gobetti 93/2, I-40129 Bologna, Italy \\
$^{5}$ \quad INAF-Osservatorio di Astrofisica e Scienza dello Spazio, via Piero Gobetti 93/3, 40129 Bologna, Italy \\
$^{6}$ \quad Center for Astrophysics | Harvard \& Smithsonian, 60 Garden St., Cambridge MA, 02138 \\
$^{7}$ \quad Black Hole Initiative at Harvard University, 20 Garden St., Cambridge MA, 02138\\
$^{8}$ \quad Eureka Scientific, 2452 Delmer Street Suite 100, Oakland, CA 94602-3017, USA\\
$^{9}$ \quad Department of Astronomy and Joint Space-Science Institute, University of Maryland, College Park, MD 20742, USA\\
$^{10}$ \quad Zentrum für Astronomie der Universit\"at Heidelberg,
 ITA, Albert-Ueberle-Str. 2, D-69120 Heidelberg, Germany\\
$^{11}$ \quad Max-Planck-Institut f\"ur Astronomie, K\"onigstuhl 17, D-69117 Heidelberg, Germany}

\abstract{ The nature and origin of supermassive black holes (SMBHs) remain an open matter of debate within the scientific community. While various theoretical scenarios have been proposed, each with specific observational signatures, the lack of sufficiently sensitive X-ray observations hinders the progress of observational tests. In this white paper, we present how AXIS  will contribute to solving this issue.
With an angular resolution of 1.5$^{\prime\prime}$ on-axis and minimal off-axis degradation, we have designed a deep survey capable of reaching flux limits in the [0.5-2] keV range of approximately 2$\times$10$^{-18}$ \fcgs~ over an area of 0.13 deg$^2$ in approximately 7 million seconds (7 Ms). Furthermore, we have planned an intermediate depth survey covering approximately 2 deg$^2$ and reaching flux limits of about 2$\times$10$^{-17}$ \fcgs ~ in order to detect a significant number of SMBHs with X-ray luminosities (L$_X$) of approximately 10$^{42}$ \lx up to z$\sim$10. These observations will enable AXIS to detect SMBHs with masses smaller than 10$^5$ \ms, assuming Eddington-limited accretion and a typical bolometric correction for Type II AGN. AXIS will provide valuable information on the seeding and population synthesis models of SMBH, allowing for more accurate constraints on their initial mass function (IMF) and accretion history from z$\sim$0-10.
To accomplish this, AXIS will leverage the unique synergy of survey telescopes such as JWST, Roman, Euclid, LSST, and the new generation of 30m class telescopes. These instruments will provide optical identification and redshift measurements, while AXIS will discover the smoking gun of nuclear activity, particularly in the case of highly obscured AGN or peculiar UV spectra as predicted and recently observed in the early Universe.
\emph{This White Paper is part of a series commissioned for the AXIS Probe Concept Mission; additional AXIS White Papers can be found at the  \href{http://axis.astro.umd.edu/}{AXIS website} with a mission overview \href{https://arxiv.org/abs/2311.00780}{here}}.}

\begin{document}

\tableofcontents
\listoffigures
\clearpage

\section{Introduction}

The origin of the very first supermassive black holes (SMBHs) in the Universe remains an open puzzle in astrophysics, attracting significant theoretical attention for the past two decades. The scarcity of observational data to address this question has added to its intrigue. Furthermore, the detection of numerous quasars powered by SMBHs that accrete matter at both high and low redshifts has underscored the importance of understanding their origins. 
It is predicted that the remnants of the first stars would have given rise to a population of relatively light black hole (BH) seeds during the early cosmic epochs.  However, the key question remains: What is the actual mass of these remnants?  

The reason for this puzzle is the detection of SMBHs with masses exceeding 10$^9$ M$_\odot$ at redshifts (z) around 7 \cite[see e.g.][]{mort,ban18,wan} and preliminary Chandra and JWST evidence of 10$^7$ M$_\odot$ SMBHs up to z$\sim$10 \cite{bogdan,Maiolino_2023}. These observations have raised an intriguing paradox. 
The conventional mechanisms for stellar remnants to grow into such massive SMBHs within the limited available time seem inadequate. This inconsistency has spurred the investigation of alternative explanations, including the presence of heavy seeds like Direct Collapse Black Holes \cite[DCBH, see e.g.][]{lod,beg}, remnants of relatively massive but lighter Population III stars (POPIII), or from BH mergers in compact clusters, or collapse of very massive stars formed through mergers of stars\cite{vol10}.

DCBHs are hypothesized to originate directly from the collapse of primordial gas clouds, bypassing the conventional process of star formation. These unique entities, formed with substantial masses from the outset, offer a compelling explanation for the swift growth of SMBHs observed during the early stages of cosmic history.
\begin{figure}[!t]
\centering
\includegraphics[width=0.80\textwidth]{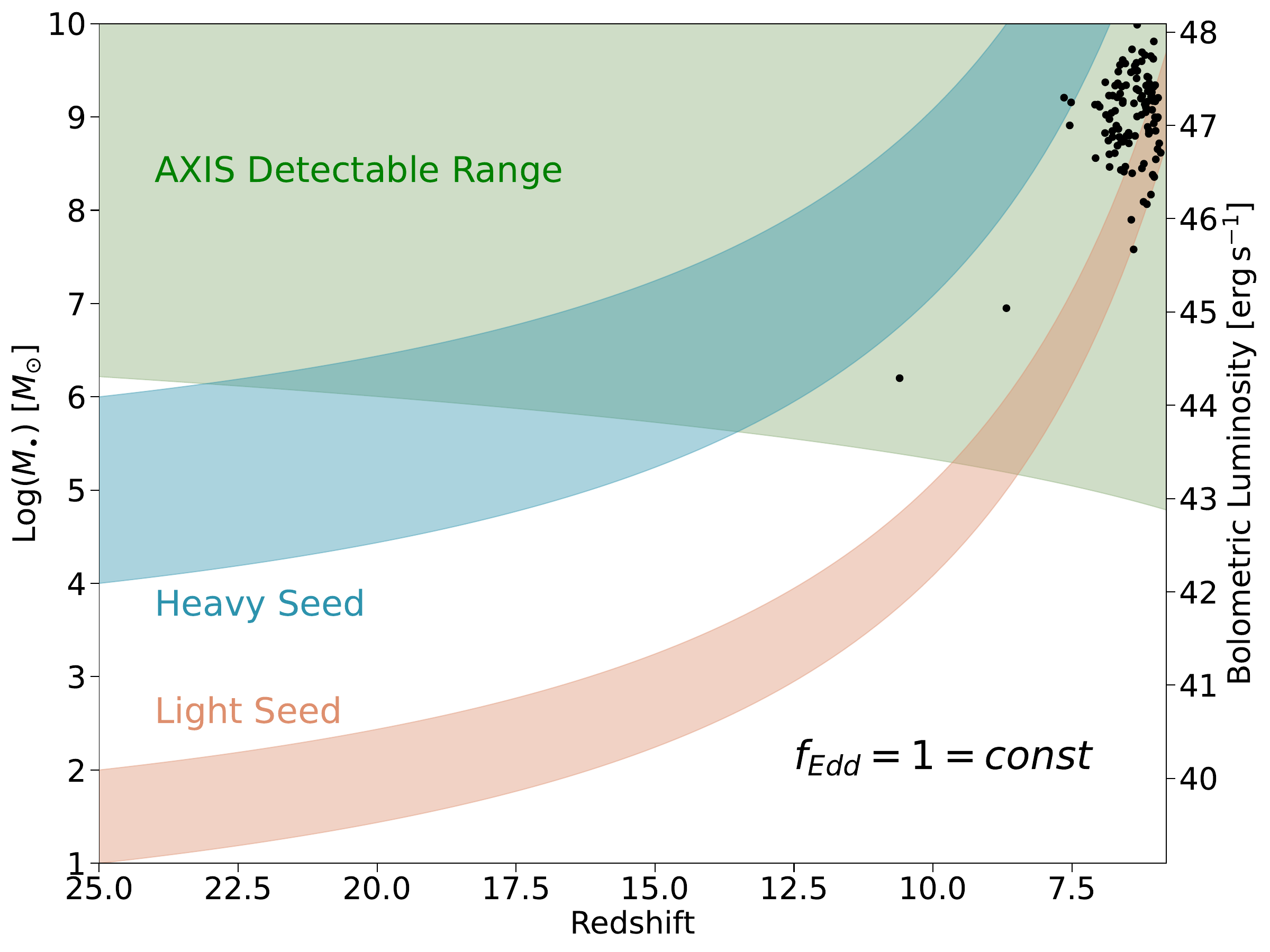}
\caption[Evolution Tracks for Black Hole Seeds]{Evolution tracks for black hole seeds (light and heavy), assuming Eddington-limited growth. The heavy seeds range between $10^4$ and $10^6 \, \rm M_\odot$ in initial mass, while the light seeds range between $10$ and $100$. Seeding occurs at a typical redshift of $z \sim 25$ \citep{BL_2001}. The black points show the masses and redshifts of known $z > 6$ quasars \citep{Fan_2022, Larson_2023, Maiolino_2023}. The X-ray bolometric correction is assumed to be 10\%. AXIS will  be able to detect all black holes within the green-shaded area. }\label{fig:growth}
\end{figure}
Likewise, remnants of POPIII stars, which are believed to have formed in environments lacking heavy elements, may retain significant masses as they undergo stellar evolution. Although these remnants are lighter than DCBHs, they still have the potential to act as seeds for the subsequent growth into supermassive entities.

In recent times, the notion of Primordial Black Holes \cite[see e.g.,][]{hawking,carr,cap22} has regained prominence as another plausible solution to the enigma of the formation of SMBH. PBHs are black holes that might have emerged directly from the extreme density fluctuations present in the early universe right after inflation. If PBHs  exist, they could offer a plentiful supply of massive seeds, facilitating the rapid growth of SMBHs within the observable timespan.

The resurrection of the PBH concept, along with the consideration of DCBHs and remnants of POPIII stars, underscores the capacity for alternative mechanisms to explain the early formation and growth of SMBHs. These ideas challenge the traditional paradigm and call for further theoretical investigations, observational constraints, and potential detection methods to gain deeper insights into the enigma surrounding the origins of supermassive black holes in the cosmos.

In Fig. \ref{fig:growth}, we present the growth of two hypothetical SMBHs that are seeded either from a "light" M$\sim$10-100 \ms seed or from a "heavy" M$\sim$10$^{4-6}$ \ms seed, in comparison with a collection of quasar detections at z$>$6. The plot is compared with the sensitivity limit of a 7 Ms AXIS observation (see below). Assuming an indicative seed collapse at z=25, the heavy seeds are capable of producing the observed high-redshift SMBH masses without requiring prolonged periods of super-Eddington accretion. In contrast, if SMBHs originated solely from light seeds, an extended phase of sustained super-Eddington growth would be necessary. However, in the case of PBHs, these assumptions are relaxed due to the considerably longer time available for black hole growth or, as suggested by certain models \cite{garcia,carr}, the possibility of PBHs forming with very large masses at the $e^+e^-$ annihilation phase transition, a few seconds after the Big-Bang.

The presence of X-ray emission is widely considered to be a definitive indicator of the existence of an actively growing supermassive black hole (SMBH) in a galaxy, observed as an Active Galactic Nucleus (AGN).
Current, highly sensitive X-ray telescopes such as Chandra and XMM-Newton have played a crucial role in advancing our understanding of the growth and evolution of AGNs up to redshifts of z$\sim$4-5. Surveys conducted with Chandra have achieved deep [0.5-2] keV flux measurements on the order of 10$^{-17}$ \fcgs, although limited to relatively small fields spanning a few square arcminutes. This limitation is due to the rapid degradation of the Point Spread Function (PSF) with increasing off-axis angles and the presence of strong vignetting. On the other hand, XMM-Newton offers a more stable but larger PSF, but having an angular resolution typically exceeding 5\arcsec on-axis limiting its sensitivity due to source confusion.
Moreover, Chandra's performance degradation has prevented it from reaching much deeper fluxes than what it has already obtained in the CDFS \cite{luo} with a reasonable time investment; therefore, the discovery space in the limiting flux-area space has stalled over the last decade. 

With a proposed on-axis angular resolution of 1.5\arcsec and an averaged Point Spread Function (PSF) Half Energy Width (HEW) of 1.6\arcsec, coupled with an effective area of 4200 cm$^2$ at 1 keV, 830 cm$^2$ at 6 keV, and a 24$^{\prime}$  diameter active field of view (i.e. $\sim$0.13 deg$^2$), AXIS \cite{2023arXiv231100780R} is poised to be a cutting-edge "X-ray survey machine" for the next decade. Benefitting from the low background provided by its low Earth orbit, AXIS will surpass Chandra by achieving depths that are one order of magnitude greater over a high-angular-resolution area that is at least 10 times larger. The net result is an effective grasp (number of detected sources per unit area) of the faintest fluxes that is at least 2 (3) orders of magnitude better than that of Chandra (XMM-Newton).

\begin{figure*}[!t]
    \includegraphics[width=\textwidth]{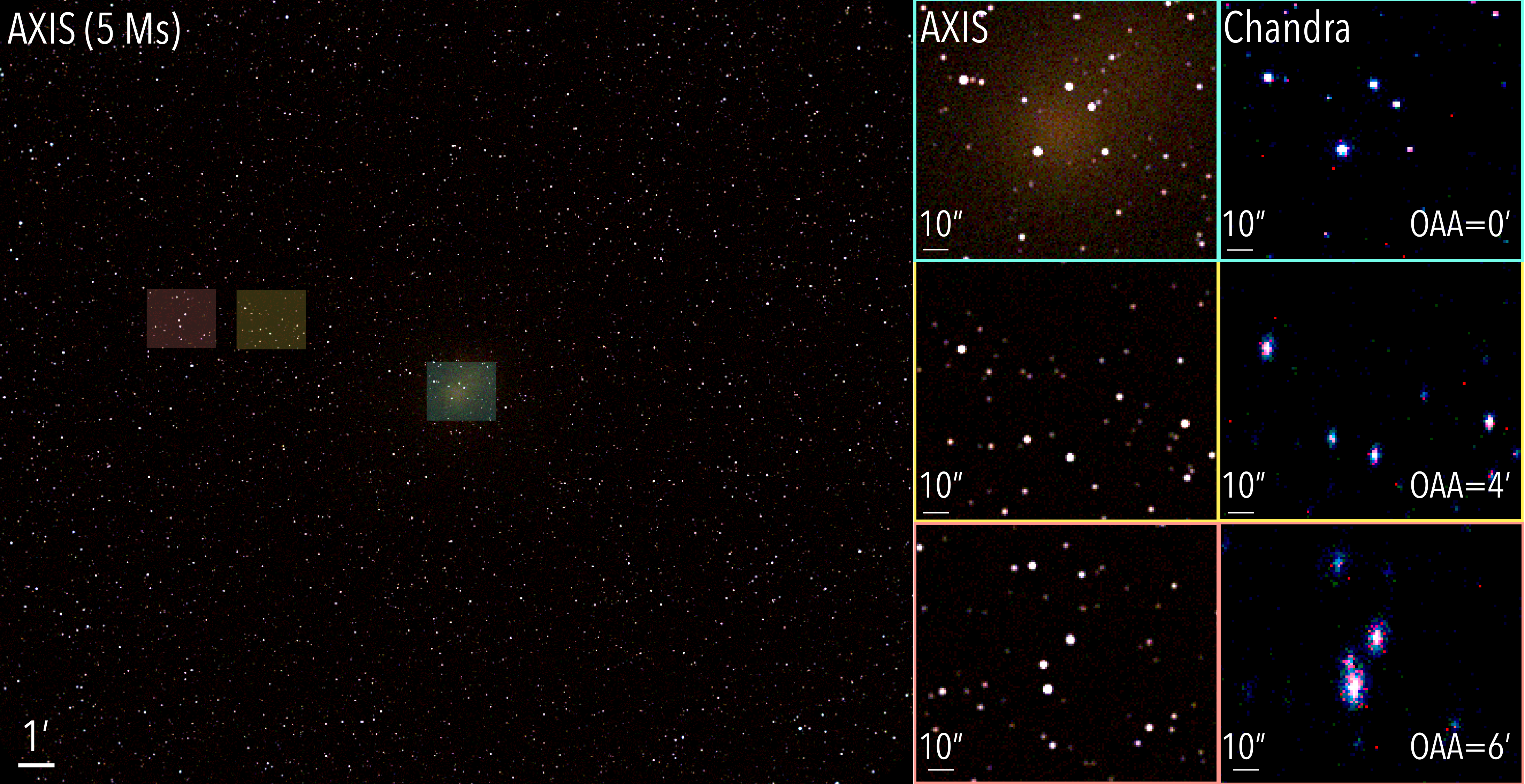}
    \caption [Comparison of AXIS and {\em Chandra} Angular Resolution]{\label{axisvsch} The large image on the left displays a simulated AXIS deep 7 Ms field. The two top right panels provide a zoomed-in view of the green area, showcasing a comparison of the on-axis performances between AXIS and Chandra. Within this area, an extended source has been included in the simulation to demonstrate how AXIS can readily detect point sources embedded within diffuse emission. The central and bottom right panels depict the yellow and red areas, respectively. The performance of AXIS optics exhibits minimal degradation in angular resolution even at large off-axis angles, whereas Chandra's point spread function (PSF) and effective area experience substantial degradation. AXIS's performance remains largely unaffected by the off-axis angle. }
\end{figure*}

In Fig. \ref{axisvsch}, we present a simulated deep survey conducted by AXIS, allowing for a comparison of its imaging capabilities with those of Chandra.

In this paper, we present how the proposed AXIS probe will revolutionize the field of high-z X-ray surveys and the quest to understand the nature and growth of SMBHs. 

\section{The Role of X-ray surveys in understanding the nature and the evolution of SMBHs.}
X-ray surveys have brought about a revolutionary understanding of the evolution of supermassive black holes (SMBHs). It is now evident that in the redshift range of approximately z$\sim$0-5, we observe a phenomenon known as "downsizing." This implies that the highest density and volume emissivity of the most rapidly growing, luminous, and massive SMBHs occur earlier in cosmic time (around z=3-4) compared to the less luminous, slowly growing, or less massive active SMBHs that dominate the Universe at z=0-1.
\subsection{Obscuration as a challenge to understand SMBH growth}
This downsizing trend is probably influenced by a combination of factors. According to the unified model of AGN, the accretion process is frequently concealed by a cold absorber, such as the torus or interstellar material. The intense radiation from luminous quasars can rapidly dissociate molecules, allowing radiation to escape rapidly. Conversely, if a SMBH accretes at a low rate or has a lower mass, the process is prolonged, and the line of sight remains obstructed for an extended period, delaying the clearing of the absorber.

The observed downsizing phenomenon could be attributed to the presence of cold absorbers, the ionizing radiation from luminous quasars, and the rate of accretion and mass of the SMBHs themselves. Understanding these mechanisms is crucial to unraveling the intricate processes that govern the growth and evolution of SMBH populations throughout cosmic history.

Indeed, recent studies have revealed an intriguing trend with regard to obscured quasars, indicating an increasing fraction of them with redshift. In particular, most (over 75\%) of the quasars at z$>$3-4 exhibit significant obscuration due to absorbing material with column densities N$_H>$10$^{23}$ \citep[e.g.,][]{peca}. These findings highlight the prevalence of heavily obscured sources at higher redshifts which are very difficult to detect with restframe optical or UV observations. 
\subsection{High angular resolution X-ray surveys to unveil the elusive nature of High-z AGN.}
Preliminary results from early observations with the James Webb Space Telescope (JWST) \cite{jwobs} have provided valuable insights into the nature of obscured sources. Specifically, at z=5, the number of type II sources is estimated to be at least three times higher than previously inferred from observations conducted by the Chandra X-ray Observatory \cite{vito}. This discrepancy can be attributed to two primary factors.

First, the volume covered by high-resolution Chandra observations is relatively limited, effectively constrained to the inner few on-axis arcminutes of the survey. Consequently, these observations do not capture the rarer, more luminous sources, potentially leading to underestimated numbers.

Secondly, these highly obscured sources are predicted to be inherently faint in X-ray emission, with [0.5-2] keV flux levels below the detection threshold of approximately 1-5$\times 10^{-17}$ \fcgs. As a result, they may go undetected in the Chandra deep X-ray surveys, further contributing to the underestimation of their abundance.

Taken together, these factors highlight the necessity of combining observations from different wavelengths, such as infrared (e.g., JWST) and X-ray (e.g., Chandra or future missions like AXIS), to obtain a comprehensive understanding of obscured sources. By leveraging the capabilities of both telescopes, we can overcome the limitations imposed by the small volume probed by Chandra and the faintness of highly obscured sources, thus providing a more accurate and complete picture of the population of obscured quasars across cosmic history.

At the same time, distinguishing highly obscured AGNs from normal star-forming galaxies based solely on Near-Mid Infrared observations is indeed challenging, if not virtually impossible. This difficulty arises due to the presence of highly degenerate features shared by both star formation processes (such as dust and narrow spectral lines) and obscured AGNs.

 This similarity in features makes it extremely challenging to untangle the true nature of a given source using only infrared observations.

To overcome this challenge and gain a more accurate understanding, a multi-wavelength approach is crucial. Combining observations from different spectral ranges, such as X-ray, optical, and infrared, allows for a more comprehensive analysis. X-ray surveys, such as those conducted by Chandra or future missions like AXIS, provide valuable information about the presence of AGN activity through the detection of X-ray emissions associated with the accretion onto supermassive black holes.

Furthermore, optical observations can provide insight into the presence of broad emission lines and other AGN signatures. By integrating data from multiple wavelengths with Near-Mid Infrared observations, a more thorough examination of sources can be achieved, leading to more accurate classification and differentiation between highly obscured AGNs and normal star-forming galaxies.

Indeed, a commonly employed technique for selecting AGNs involves the use of rest-frame (observed NIR at high-z) optical spectroscopy, typically employing the BPT diagram. However, early results from JWST indicate that this approach may not be as effective at high redshifts, as the spectra of AGNs at these cosmic epochs differ from those commonly observed in the local Universe.

For example, a recent study by \cite{Maiolino_2023} presents the case of an AGN at z$\sim$10.6, where an unusually bright Ne [IV] line is observed. This emission line is atypical for AGNs \cite{terao}, highlighting the distinct nature of high-redshift AGN spectra. Additionally, a recent survey of high-redshift broad-line AGNs using JWST has revealed a lack of [NII] emission \cite{harikane}. As a consequence, these AGNs fall within the parameter space characterized by high [OIII]/H${\beta}$ ratios and low [NII]/H${\alpha}$ ratios, similar to galaxies at z$\gtrsim$4 without broad-line emission. This finding suggests that the low metallicity of their host galaxies prevents their differentiation from regular star-forming galaxies using the BPT diagram.

However, it is worth noting that at higher luminosities, this phenomenon seems to be less pronounced. Hence, low-mass SMBHs (on the order of 10$^{6-7}$ \ms) growing in low-metallicity, high-column-density galaxies necessitate confirmation in the X-ray band to be definitively labeled as AGNs. X-ray observations, such as those conducted by AXIS, can play a crucial role in confirming the presence of AGN activity in these sources.
\subsection{The power of combining AXIS and JWST surveys}
Therefore, the distinctive spectroscopic features observed in high-redshift AGNs, coupled with their occurrence in low-metallicity environments, pose challenges for their identification using traditional methods such as the BPT diagram. Complementary observations in the X-ray band are essential for providing definitive confirmation of AGN activity in these specific scenarios.

Indeed, the combination of JWST and AXIS is of utmost importance in comprehending the nature of galaxies hosting the first AGNs, as well as the coevolution of supermassive black holes (SMBHs) and galaxies.

\begin{figure}
    \centering
    \includegraphics[scale=0.7]{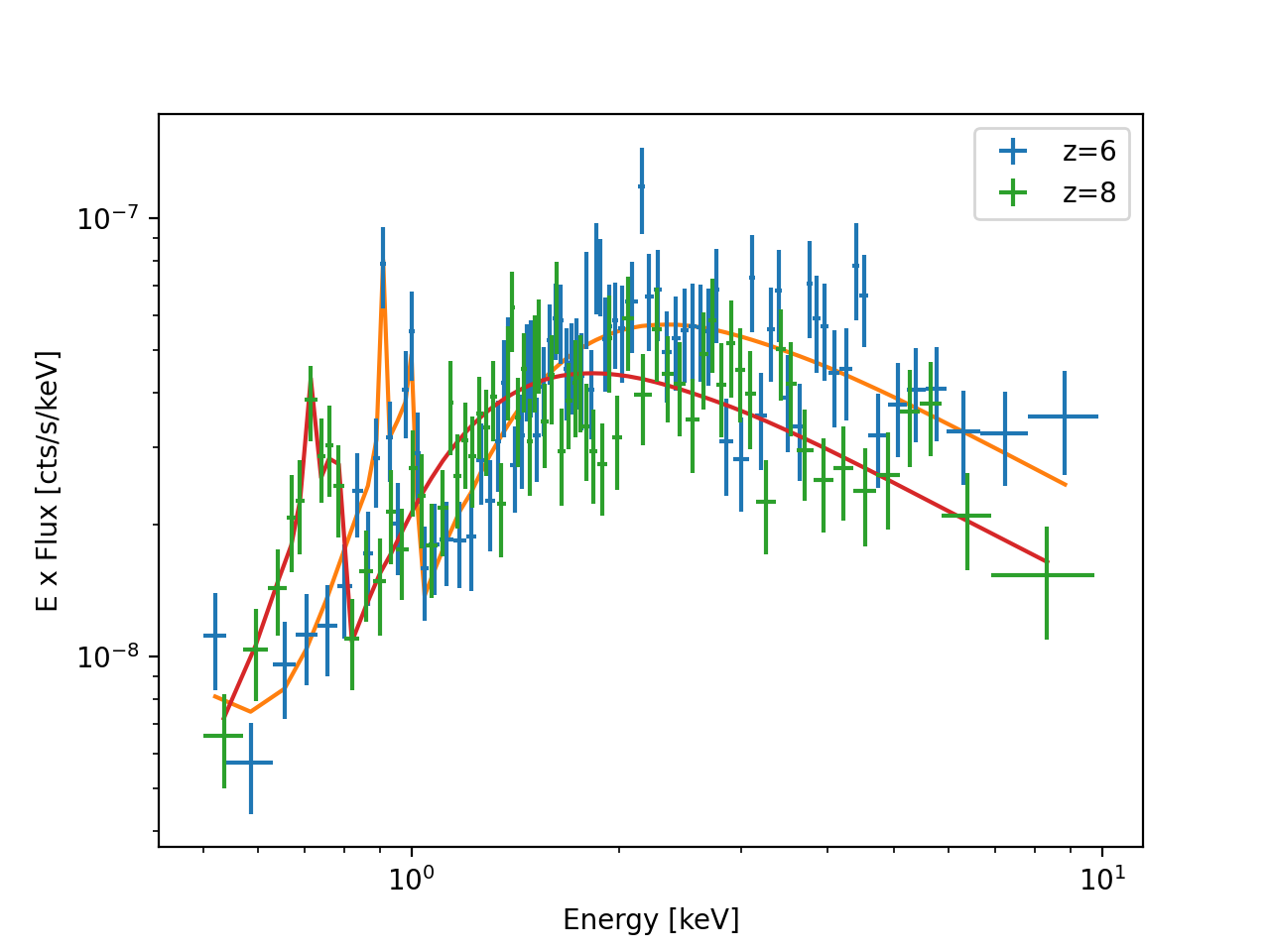}
    \caption[AXIS Simulated X-ray Spectrum of High-z Highly Obscured AGN]{
    \label{fig:ct} Simulation of a 7 Ms observation of a Compton Thick AGN with an intrinsic luminosity of L$_X$=2$\times$10$^{44}$ \lx at redshifts around 6 and 8. AXIS will detect hundreds of counts, enabling the measurement of column density (N$_H$) at a significance level of 5$\sigma$. The continuous lines represent the best fits obtained using the XSPEC {\em MyTorus} model. This level of statistical accuracy is comparable to that achieved by NuSTAR in studying Compton Thick AGN in the local Universe.}
\end{figure}

Soft X-ray surveys play a crucial role in this endeavor, as obscured AGNs, including Compton-thick sources, exhibit peak emissivity in the 10-20 keV energy range. At redshifts z$>$6, this peak is shifted into the [0.5-2] keV energy range, precisely where focusing X-ray telescopes like AXIS possess their maximum effective area, as depicted in Figure \ref{fig:ct}. 

This advantageous alignment allows AXIS to detect sources that JWST may not be able to identify as AGNs because of the relatively unbiased nature of X-ray selection. By conducting soft X-ray surveys, AXIS can identify and characterize obscured AGNs even in cases where they may be missed or misclassified based on other observational criteria. This unbiased selection is essential for a comprehensive understanding of the AGN population and its connection to the early Universe.

The synergistic combination of JWST and an X-ray telescope like AXIS enables a holistic view of galaxy evolution, shedding light on the intricate interplay between SMBHs and their host galaxies. By harnessing the power of multi-wavelength observations, we can explore the coevolutionary processes that shape these systems throughout cosmic history.

To study SMBHs at redshifts z=6-10, it is crucial to have an observatory capable of detecting sources with intrinsic X-ray luminosities L$_X>$10$^{42}$ \lx, and log N(H)>23 as most of the accretion in this epoch is expected to be obscured. The   X-ray luminosities of highly obscured, Eddington limited, M$\gtrsim$7$\times$10$^5$ \ms, sources correspond to fluxes of approximately 1-4$\times$10$^{-18}$ \fcgs in the [0.5-2] keV energy band. Such sensitivity levels are necessary to probe the presence and properties of rapidly growing BHs in the high-reshift universe. 

AXIS, with its designed capabilities and sensitivity, is precisely tailored to achieve these required limits. Combined with the survey strategy outlined below, AXIS has the potential to serendipitously detect hundreds of SMBHs at redshifts z=6-10 and even extend the search to z$\sim$12 in the case of pointed observations targeting pre-selected objects.
\subsection{SMBH and host galaxy coevolution over cosmic time with AXIS}
The combination of AXIS's sensitivity to faint X-ray sources and its survey strategy will significantly advance our understanding of the early Universe, shedding light on the formation and evolution of SMBHs during these cosmic epochs. By enabling the detection of a large number of SMBHs at high redshifts, AXIS will contribute to our knowledge of the coevolution of galaxies and their central supermassive black holes and its early onset, providing crucial insights into the early stages of cosmic structure formation.

Studying the growth of supermassive black holes (SMBHs) from the earliest epochs of the universe with redshifts greater than 6, through to the peak of star formation at redshifts 1-2, is crucial for unraveling the intertwined evolution of SMBHs and galaxies. During this period, most stars in the universe formed, making it a pivotal phase in cosmic history.

Recent advances in our understanding of active galactic nuclei (AGNs) have revealed that they are not constant sources of light. Instead, they exhibit variability in their Eddington ratios on relatively short timescales, typically less than 10$^{5-6}$ years \citep{hickox14}. As a result, it is essential to explore the full range of accretion rates and luminosities across the entire cosmic timeline in order to comprehensively characterize the AGN population.

Additionally, based on new JWST data of galactic systems in the redshift range $z=4-7$, with both SMBH and stellar mass measurements available, \cite{Pacucci_2023_overmassive} recently found that SMBH masses are $10-100$ times higher than what the host's stellar mass would suggest, based on the local scaling relations. This study indicates a significant number of yet-to-be-detected SMBHs in current JWST surveys. They are still undetected because (i) they may be inactive (i.e., duty cycles $1\%-10\%$), (ii) the host overshines the AGN, or (iii) the AGN is obscured and not immediately recognizable by line diagnostics. Next-generation X-ray missions, such as AXIS, could then play a fundamental role in uncovering this, thus far, hidden population of sources.

This approach has been successfully implemented by \cite{ananna22,ananna22b} at redshift z$\sim$0, and future studies extending this methodology to higher redshifts will be crucial for obtaining a complete understanding of SMBH growth and its coevolution with host galaxies. By investigating the interaction between the growth of large-scale structures and the accretion rates on SMBHs, profound advances can be made in our understanding of the AGN-galaxy coevolution \citep[e.g.,][]{powell18,powell20}.

By examining the intricate relationship between SMBH growth, star formation, and structure formation across different cosmic epochs, we can gain transformative insights into the processes that govern the coevolution of AGNs and their host galaxies. This comprehensive understanding will contribute to our broader knowledge of galaxy evolution and the formation of the cosmic structures that we observe today.

\section{AXIS Pillar 1: Determining the Nature of SMBH Seeds with AXIS at z = 6-10 XLF}

The growth of SMBHs by accretion and mergers gradually erases seeding information.  However, the closer to the seeding epoch that we can probe AGN, the closer we get to their initial masses and host galaxy relationships.  The epoch of $z \gtrsim 7$ is a special time: only at these extreme redshifts do quasars require seeds to have grown continuously near the Eddington limit to acquire their observed masses.  This places interesting joint constraints on the astrophysics of SMBH seeding and accretion, which grow more and more informative as redshift increases.

Using a semi-analytical model (SAM) that integrated the SMBH assembly from $z=20$ to $z=0$, \citet{Ricarte&Natarajan2018} simultaneously studied gravitational wave signals, luminosity functions, and occupation fractions to search for signatures that could be used to discriminate seeding models.  This model contrasted rare and ``heavy'' seeds from direct collapse with abundant and low-mass light seeds from the remnants of Population III stars.  Using empirical relations, these seeds grow at a rate capped at the Eddington limit with a duty cycle modulated by major halo mergers.  Interestingly, the luminosity functions of these two cases are identical for $z \lesssim 8$and down to L$_X\sim$10$^{42}$ \lx, where AGN behavior was determined almost entirely by SMBH-galaxy co-evolution.  However, the two cases diverge at higher redshift, nearer to the seeding epoch, where the Eddington limit leads to very different high-to-moderate luminosity functions.  We note that there are three major factors that determine the luminosity function at this epoch: BH masses, occupation fractions, and accretion rates.  In combination with other facilities, AXIS will enable us to disentangle these effects.

\begin{figure}
    \centering
    \includegraphics[width=\textwidth]{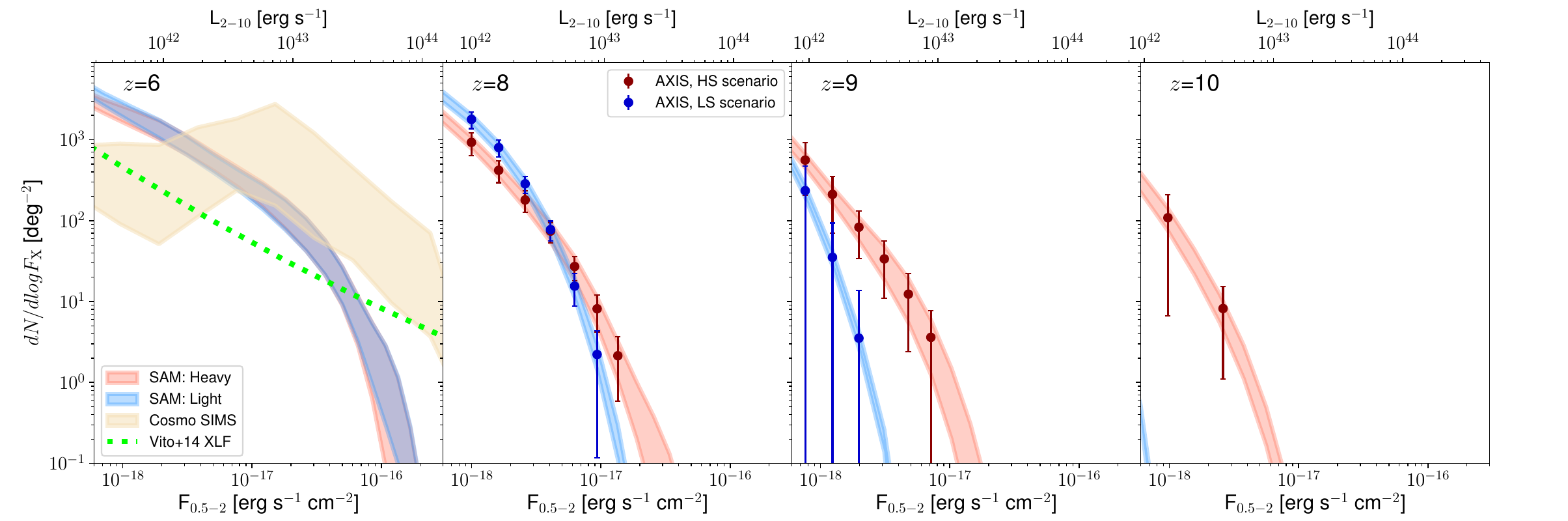}
    \caption[The z$>$6 XLF of Light and Heavy Seeds AGN with AXIS Surveys]{The expected source counts from both semi-analytical models (SAMs) and cosmological simulations are shown in the figure. The red and blue curves represent heavy and light seeding scenarios, respectively, from the SAM developed by \citet{Ricarte&Natarajan2018} (see text for more details). These curves diverge at redshifts above 9, highlighting the different predictions for these scenarios. The cosmological simulations compiled by \citet{2022MNRAS.509.3015H} exhibit a wide range of luminosity functions, underscoring the need for additional high-redshift constraints to better constrain the models. Additionally, at redshift 6, the 3$\sigma$ upper uncertainty on the number counts from the X-ray luminosity function by \citet{vito14} is shown for reference. This uncertainty is extrapolated from observational evidence at redshift 4, providing additional context for the expected source counts at high redshifts.}
    \label{fig:seeding}
\end{figure}

In \autoref{fig:seeding}, we plot the source counts predicted from \citet{Ricarte&Natarajan2018} SAM (heavy seeds in red, light seeds in blue). We add in orange a compilation of LF obtained from the large-scale cosmological hydrodynamical simulations of the field \citep{2022MNRAS.509.3015H}. The scatter at fixed Lx reflects the impact of different seeding, BH growth, and feedback modeling in these simulations.\footnote{Values are collected in Habouzit et al. in prep. from {\sc Horizon-AGN} \citep{2016MNRAS.463.3948D}, {\sc Illustris} \citep{2014MNRAS.445..175G}, {\sc EAGLE} \citep{2015MNRAS.446..521S}, {\sc TNG100} and {\sc TNG300} \citep{2018MNRAS.473.4077P}, {\sc SIMBA} \citep{2019MNRAS.486.2827D}, and {\sc Astrid} \citep{2022MNRAS.513..670N}.}  The points with errorbars represent the combination of the proposed ``Deep'' and ``Intermediate'' survey strategies, assuming the SAM points as true.  SAM bolometric luminosity functions have been converted to 2-10 keV assuming bolometric corrections of \citet{Duras+2020} for Type II AGN and have been additionally suppressed by an ad hoc factor of 10, assuming an extreme Compton-thick fraction.  Despite all being calibrated to eventually yield the correct local scaling relations between SMBH mass and galaxy mass, the cosmological simulations span a wide range, implying that additional constraints at this epoch would help constrain the astrophysics of this epoch.  As discussed above, the two SAM curves begin to diverge at $z \gtrsim 8$, due to the assumed Eddington limit.  Since heavy seeds were initialized at higher masses, they are able to produce AGN with higher luminosity at high redshifts than their light seed counterparts.  A measurement of the evolution of the luminosity function during this epoch would place interesting joint constraints on seeding and accretion. 
The main goal of AXIS will be to unambiguously detect SMBH at z$>$7  to derive the XLF to inform and constrain seeding models. This will only be possible by performing an X-ray survey with AXIS in the region where NIR measurements with JWST and/or Roman exist or follow-ups are possible to properly characterize the host galaxies.


So far extrapolations from population synthesis models differ of up to two orders of magnitude from theoretical models, but even the deepest Chandra surveys are not able to discriminate among them at $z\gtrsim6$. In fact, regardless of the assumed model, $<<1$ AGN is expected to be detected with Chandra at $z>6$ with flux $<10^{-17}\,\mathrm{erg/,cm^{-2}\,s^{-1}}$, since only $\approx10\,\mathrm{arcmin^2}$ in the 7 Ms Chandra Deep Field-South are sensitive to such emission level. 

\subsection{Accretion mode and signatures of the first SMBHs}
To date, we detected more than $300$ quasars at $z>5.9$, with tens of them already characterized by mass estimates \citep{Fan_2022}. Several confirmed detections at $z > 7$, including some recent striking observations of candidate AGN by JWST at $z\sim$6-12 \citep{Larson_2023, Maiolino_2023,epochs}, provide us with a test bed for our models of black hole growth.

Seeds, light and heavy, are predicted to form in the redshift range $z \sim 20-30$ \citep{BL_2001}. Because the detection redshifts of quasars are increasing, the time between seed formation and SMBH detection is decreasing \cite[see, e.g.][]{Euclid_2019, Fan_2022}. The decreasing time between seeding and detection shrinks the parameter space of seed properties \citep{Pacucci_2022}: initial mass, average Eddington ratio, duty cycle, and radiative efficiency. Consequently, detecting SMBHs at very high redshift allows us to pinpoint the properties of black hole seeds in the most efficient way. For example, detecting a SMBH with mass $\sim 10^{10} \, M_\odot$ by $z \sim 9$ would exclude the entire parameter space available for light seeds and dramatically reduce the one for heavy seeds \citep{Pacucci_2022}.
The detection of a more moderate case (e.g., GNz11, if confirmed, a $\sim 10^{7} \, M_\odot$ at $z \sim 10$) would still provide important constraints on the seeding mechanisms, especially if the (instantaneous) Eddington ratio is constrained. For example, \cite{Pacucci_2022} shows that, in this case, the marginal distributions for the seed mass for eight quasars already detected at $z > 7$ can already be constrained down to 2 orders of magnitude in mass. Modeling of some of these quasars already shows a preference for heavy seeding scenarios.

Detecting accreting SMBHs at $z>7-8$ is one of the main goals of AXIS. With its surveys, AXIS will be able to detect low-luminosity AGN up to the highest redshifts probed spectroscopically with JWST (see Fig. \ref{fig:growth}). By detecting X-ray emission from these sources, a first estimate of their black hole mass will be available, assuming accretion at the Eddington rate. This first analysis can be complemented by follow-up observations with JWST, either with photometry or, for sources that are sufficiently bright, with spectroscopy. This additional information will constrain the mass and the redshift of the source. Hence, improved modeling with those as priors will further pinpoint the initial seed mass and the average growth rate (i.e., sub-Eddington vs. super-Eddington) to a precision that is unreachable today with current facilities.

In the electromagnetic realm, signatures of heavy vs. light seeding can be grouped into two categories: seed-related and host-related.
Seed-related signatures are generated from the black hole itself. Early studies (see, e.g., \cite{Pacucci_2015, Valiante_2018, Ricarte_2018}) show that heavy seeding will lead to steep, red infrared spectra and a bell-shaped X-ray emission peaked at $\sim 1$ keV rest-frame. While current X-ray facilities, such as Chandra, are able to detect extremely massive, unobscured seeds, next-generation facilities are urgently needed to detect broader, fainter, and obscured populations. With a peak emission, at $\sim 1$ keV, 1-2 orders of magnitude fainter, obscured populations of seeds will remain unobserved until next-generation instruments are operative \citep{Pacucci_BAAS}. 
AXIS will thus play a crucial role in detecting the Compton-thick population of heavy black hole seeds. The obscured fraction of seeds is expected to represent the norm at such high redshifts \citep{Pacucci_2017}.

Host-related signatures, such as the black hole-to-stellar mass ratio, are essential in discriminating between light- and heavy-seeding models. In the heavy seed case, the ratio is expected to remain of order unity for several Myrs after the seeding process (see, e.g., \cite{Visbal_2018}). On the contrary, in the light seeding case, the ratio is expected to be much lower and progressively stabilize around the local ratio of $\sim 10^{-4}$ over some Gyrs. Assuming typical effective radii nowadays detected for $z \sim 10$ hosts, JWST will be able to detect their infrared light and constrain their stellar mass \cite{Naidu_2022, Adams_2023}.
As X-ray observations are crucial both for studying the spectral energy distribution of the X-ray emission and for estimating the mass of the SMBH, AXIS will become the facility of reference to investigate both populations of black hole seeds.

The X-ray spectra of AGN are also a direct tool for studying the accretion mechanism and the properties of the accretion discs. In the redshift range $0<z<6$ \cite{vitospec} reports a lack of substantial evolution of the inner accretion-disk and hot-corona structure in QSOs. Surprisingly, recent results from \citep{zappa} show that a sample of very luminous broad-line AGN z$>$6 presents an unusually steep X-ray spectrum with spectral index $\Gamma\sim$2.4 compared to average $\Gamma\sim$1.9 observed at z$<6$. This points to a different physics of the disc and the corona of the AGN at  high-z which, combined with the peculiar requirements of fast growth of the SMBH at high makes the study of the physics of accretion at that epoch even more intriguing. AXIS will be able to study the spectra of hundreds of AGN at z$>$6 allowing the study of accretion physics over a much wider range of luminosities and accretion rates. Another suggestion of \cite{zappa}, which cannot be currently tested because of the low count statistics, is that these sources could instead present an unusual extremely low-energy cutoff and a typical $\Gamma$=1.9 spectral index.  AXIS with its very large effective area will be able to produce much better spectra than XMM-Newton at high-z and further test the accretion physics of AGN in the early Universe.

\section{Assembly of black holes with large-scale structures
as traced by high-z AGN clustering}

A complementary way to study the growth of supermassive black holes is by connecting them to their larger-scale dark matter environments, probed by the spatial clustering of AGN. Together with the luminosity function, AGN clustering constrains the relationship between black holes and their host dark matter halos and infers any preferred cosmic environments for AGN activity. Since the growth and evolution of halos is well understood, clustering can be used to statistically connect SMBHs to their progenitors in the earlier universe, which would infer how certain populations evolve over time.

The area and depth of the intermediate AXIS survey (with redshift measurements from future multiwavelength facilities; see Section 5.2.2) would allow the uncertainties of AGN clustering at $z\sim3$ to be at the 10\% level or better on both the large {\it and small} scales ($0.1-10$ Mpc/h). This is an order-of-magnitude better than the Athena ESA concept mission, and unfeasible with current instruments. Cross-correlations with galaxies at intermediate redshifts ($z\sim 1.5$) are predicted to provide uncertainties less than 5\%. The unprecedented clustering statistics at these epochs (especially on 1-halo scales; $<1$ Mpc/h) would allow for measurements in several bins of AGN luminosity, which would provide powerful constraints on AGN fueling and feedback mechanisms at the peak of SMBH growth (e.g., determining whether or not the majority of AGN are triggered by galaxy mergers, as well as constraining the environmental dependence of the distribution of Eddington ratios).

Finally, JWST will dramatically improve the characterization of the close environments around high redshift ($z>6$) AGN, via observations of galaxy fields around AXIS AGN of varying luminosities. This will determine whether or not high-z AGNs reside in overdensities for a wide range of accretion rates and mass scales. Although theory predicts that the most massive black holes at high redshift reside in the progenitors of today's massive clusters, observations have so far led to conflicting conclusions, most likely due to the limited HST sensitivity. JWST will detect high-z galaxies much more efficiently, and together with the wide range of AGN luminosities from the AXIS surveys, this long-standing problem will be solved.

Moreover, the clustering statistics of over-(or under-)massive SMBHs at high-z will shed light on their seeding mechanisms, as those that originated from DCBHs are predicted to be significantly more clustered than SMBHs formed via other mechanisms. This is because the Lyman-Werner radiation from one DMBH would likely trigger additional DCBHs within the same region, resulting in highly clustered AGN statistics in the early universe. Therefore, the clustering amplitudes of AXIS AGN will provide powerful additional constraints on such SMBH formation models.
\section{AXIS Survey of The First AGN}
Surveys with the AXIS probe have been first simulated and discussed in \citet{Marchesi20}, to which we refer for a more detailed explanation of the simulations. 
Both the \citet{Marchesi20} simulations and the ones used to obtain the numbers reported in this paper have been performed using the Monte Carlo code Simulation of X-ray Telescopes \citep[hereafter \texttt{SIXTE},][]{dauser19}. This software allows one to simulate an observation with an X-ray telescope by creating a photon list, which includes the arrival time, energy, and position of each photon based on the simulated telescope setup (i.e., effective area, field of view, point spread function, vignetting, read-out properties, redistribution matrix). In doing so, \texttt{SIXTE} generates an impact list that contains the energy and arrival time of each photon, as well as its position on the detector. The final event file is obtained from this intermediate list, reprocessed to take into account the simulated detector read-out properties and redistribution matrix file.

The input catalog we used for our simulations is based on the \citet{gilli07} AGN population synthesis model. AGN have been simulated down to a 0.5--2\,keV luminosity L$_{0.5-2}$=10$^{40}$\,erg\,s$^{-1}$ and up to redshift $z=$3. The mocks have been checked to ensure the correct reproduction of the trends with luminosity, redshift, and column density of the AGN densities as a function of luminosity, redshift, and column density, and are in close agreement with the observational evidence at all fluxes sampled by current X-ray surveys. In the high-redshift regime (i.e., at $z>$3, where the AGN space density starts declining), we use a separate mock catalog, built from the \citet{vito14} $z>$3 AGN luminosity function, which nicely describes the observational evidence from the deepest X-ray surveys currently available up to redshift $z\sim$5 \citep[e.g.,][]{vito18}.
Finally, neither AGN nor host galaxy clustering are included in the mock generation. The catalogs we used are available online at \url{http://cxb.oas.inaf.it/mock.html} in FITS format and ready to be used within \texttt{SIXTE}. 

\subsection{The AXIS wedding Cake Survey}
Following a standard approach extensively used in past X-ray surveys, AXIS plans to use a so-called ``Wedding cake'' strategy to perform its extragalactic surveys, as follows.

\begin{enumerate}
    \item A deep, pencil-beam survey, that is, a 7\,Ms observation of a single AXIS pointing ($\sim$0.13\,deg$^2$).
    \item An intermediate--area, and intermediate--depth survey, which would cover 2\,deg$^2$ with an uniform, 360\,ks exposure, for an overall time request of 6\,Ms. 
    \item Finally, while no wide--area survey is currently planned, it would be possible to obtain an AXIS Serendipitous field by combining Guest Observer observations. Assuming 20\,Ms of GO non-Galactic plane time, with a median of 50\,ks per pointing, it would be possible to cover 50\,deg$^2$ with a sensitivity $\sim$10$^{-16}$\,erg\,s$^{-1}$\,cm$^{-2}$.
\end{enumerate}

We report in Table~\ref{tab:survey_layout} a summary of the properties of these surveys: the number of detections and flux limits have been obtained from end-to-end simulations with the \texttt{SIXTE} tool described in the previous section.  Note that in this table, we report only the expected number of sources at z$<$6, as the model used to generate the mocks employs a simple power-law extrapolation of z$\sim$3-4 XLF. In the AXIS Deep Field, the 3\,$\sigma$ upper boundary of the \citet{vito14} XLF predicts that one should detect $\sim$25 AGN in the redshift range $z$=[6--7]; in the Intermediate Field, instead, in the same redshift range one should detect $\sim$80 sources. In the same redshift range, the SAM heavy seed scenario predicts $\sim$120 detections in the Deep Field and $\sim$240 in the Intermediate Field, while the SAM light seed scenario 120 detections in the Deep Field and $\sim$230 in the Intermediate one. At higher redshifts, as shown in Figure~\ref{fig:seeding}, the two SAM models diverge more significantly, and the expected number of AXIS detections is much larger. For example, at $z$=9 the SAM Heavy seed scenario predicts 10--12 detections in the two AXIS fields, while the SAM Light seed scenario predicts only 1--2 detections.


\begingroup
\renewcommand*{\arraystretch}{1.5}
\begin{table*}
\centering
\scalebox{0.88}{
\vspace{.1cm}
  \begin{tabular}{cccccccc}
       \hline
       \hline      
Survey & Area & Tile exposure & Total exposure & \multicolumn{3}{c}{Flux limit (0.5--2\,keV)} & Number of AGN detections\\
       & deg$^2$ & ks         & Ms             & \multicolumn{3}{c}{erg s$^{-1}$ cm$^{-2}$}   & \\
       &         &            &                & 1\,\% & 20\,\% & 80\,\%                      & \\     
       \hline
Deep &  0.13 &  5800 & 7 & 4.5$\times$10$^{-19}$ & 1.9$\times$10$^{-18}$ & 4.3$\times$10$^{-18}$ & 2800 \\ 
\hline
Intermediate     &  2 & 360 & 6 & 4.5$\times$10$^{-18}$ & 1.1$\times$10$^{-17}$ & 2.6$\times$10$^{-17}$ & 21000 \\ 
    \hline
    \hline
	\vspace{0.02cm}
\end{tabular}}
	\caption{\normalsize Properties of three reference AXIS surveys simulated in this work. The flux limits are calculated at 1\,\%, 20\,\%, and 80\,\% of the covered field, and correspond to the fluxes at which which 1\,\%, 20\,\%, and 80\,\% of the input sources are detected. The number of detections is computed in the 0.5--7\,keV band.
	}
\label{tab:survey_layout}
\end{table*}
\endgroup
\begin{figure}[h]
    \centering
    \includegraphics[width=0.8\textwidth]{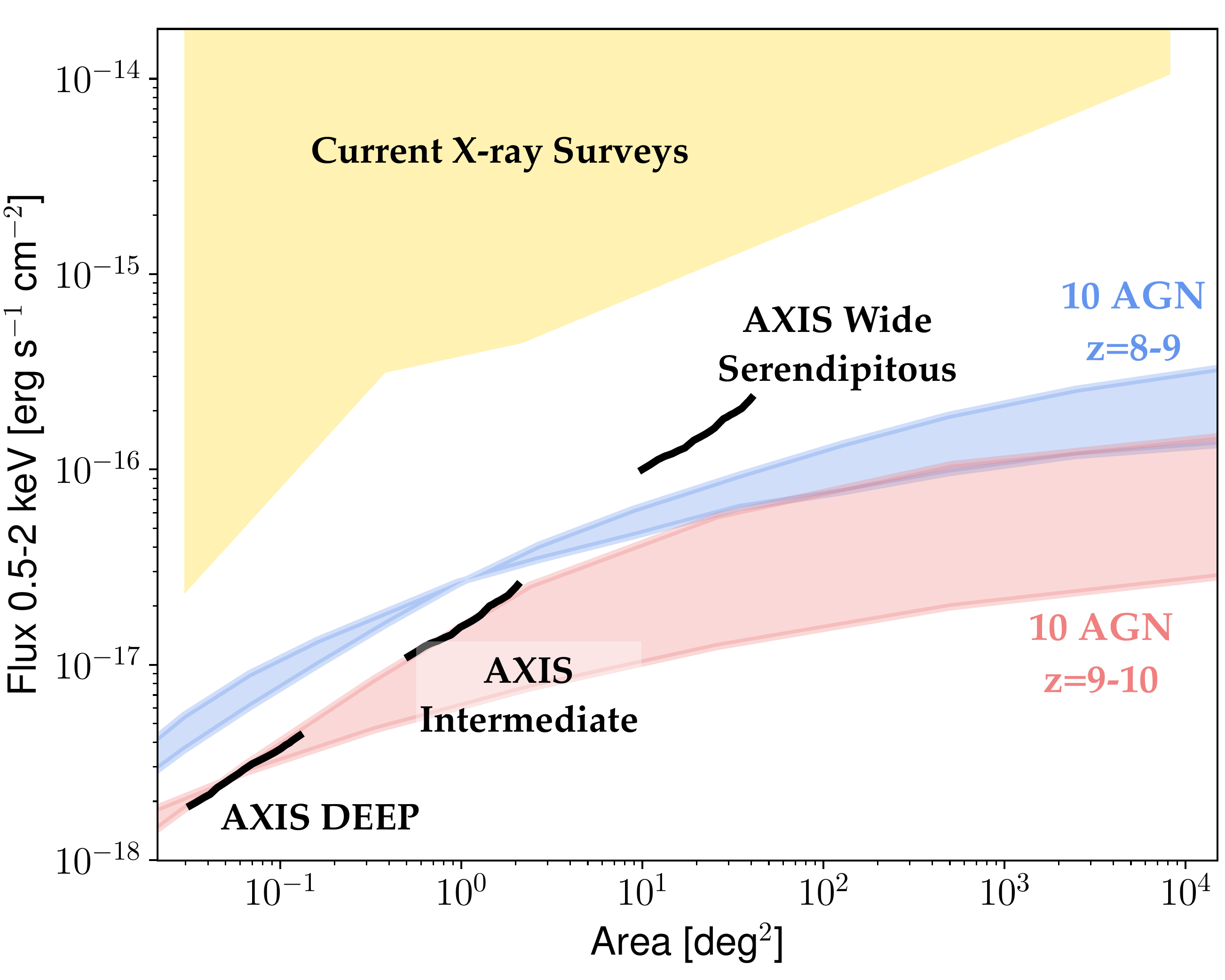}
    \caption[AXIS Survey Sensitivity]{ The figure displays the 0.5--2,keV area-flux curves for the AXIS Deep, Intermediate, and Wide Serendipitous surveys, as outlined in this paper. The plotted lines represent the sensitivity curves of the surveys in the 0.5--2,keV range, covering an area range of 20\% to 80\% of the survey coverage. The light blue and pink areas indicate the parameter space covered by the SAM heavy- and light-black hole (BH) seed models, respectively. These areas represent the predicted range of AGN detections in the redshift ranges of $z=[8-9]$ and $z=[9-10]$, assuming the detection of 10 AGN in each range. As discussed in the text, the AXIS Deep and Intermediate surveys have the potential to detect a sufficient number of sources at $z\sim8-10$, resulting in a significant reduction in the prediction range of the current SAM models. For comparison, the figure also includes a yellow area representing the parameter space covered by current X-ray surveys using \textit{Chandra}, XMM-\textit{Newton}, and eROSITA. This provides context for the coverage and capabilities of these existing surveys compared to the proposed AXIS surveys.
}
    \label{fig:enter-label}
\end{figure}
\subsection{Synergies}

AXIS is proposed to fly during an unprecedented epoch for extragalactic surveys. In the early 2030s, NASA and ESA will boast an exceptional fleet of survey satellites, while powerful ground-based telescopes will observe the first light of the universe. JWST, having completed its first decade of operation, will have already amassed a wealth of near-infrared to mid-infrared photometric and spectroscopic data from various deep survey fields. Roman and Euclid, covering a significant portion of the extragalactic sky in the visible and near-infrared bands, will contribute to the comprehensive view of the universe. Meanwhile, LSST-Rubin will observe the extragalactic sky with high cadence, capturing transient and variable phenomena, while the ELT (Extremely Large Telescope) will make its debut as a cutting-edge 30-meter class telescope. Additionally, the SKA (Square Kilometre Array) will provide a groundbreaking perspective of the radio sky.

Within this vibrant observational landscape, AXIS will serve as the ideal X-ray counterpart, completing the panchromatic view of the extragalactic universe. By harnessing its capabilities in X-ray observations, AXIS will play a vital role in unequivocally detecting accreting objects, such as AGN and other X-ray-emitting sources. Its synergistic partnership with the aforementioned observatories and telescopes will allow for a comprehensive understanding of the properties, dynamics, and coevolution of objects across the electromagnetic spectrum. This collaborative effort will significantly advance our understanding of the extragalactic universe and provide a holistic view of the diverse phenomena that shape our cosmic landscape.

\subsubsection{Optical Identifications}

While X-ray emission serves as a clear indicator of accretion in galaxies, a mere detection is not sufficient to obtain informative insights. However, due to the high angular resolution of AXIS, we will have the ability to achieve optical identification for more than 95\% of the detected sources, as estimated by the COSMOS-Legacy survey \cite{clegacy}. To accomplish this, we will rely on NIR source catalogs from observatories such as HST, Subaru, JWST, Roman, Euclid, and potentially the ELT.

Optical identification involves more than just positional cross-matching, as obtaining precise source centroids is crucial. It also requires leveraging prior knowledge of the colors and magnitude distribution of potential counterparts. To facilitate these complex optical identifications, various tools have been developed, including LYR and N-WAY. These tools use well-tested Bayesian approaches to efficiently perform the intricate process of optical identification \cite{nway2,peca,nway}.

By combining the X-ray detections from AXIS with accurate optical identifications, we can establish robust associations between the X-ray sources and their optical counterparts. This multi-wavelength approach not only provides valuable information about the nature of the accreting objects, but also enables detailed studies of their properties, including redshifts, spectral features, and host galaxy characteristics. The combination of X-ray and optical data, supported by advanced identification techniques, will significantly enhance our understanding of the connection between accretion processes and the larger astrophysical context.

\subsubsection{Photometric and Spectroscopic Redshifts}

To accurately measure the luminosity function of high-redshift AGNs, obtaining their redshift information is of fundamental importance. The survey fields targeted by AXIS will be strategically chosen in highly investigated areas of the sky, including regions such as COSMOS, Chandra Deep Fields, JWST deep Fields, and areas covered by either Roman or Euclid, or both.

The counterparts of high-redshift AGNs are expected to be extremely faint, with magnitudes on the order of m$_{AB-3.6 \mu m}\sim$26-29. Preliminary results from JWST indicate that the NIR magnitudes of the first AGN candidates identified at redshifts z$\sim$7-10 fall within the range of 25-26. By cross-matching these sources with AXIS data, they can be easily identified, especially considering that the typical flux limit of deep JWST surveys is around 29 (AB mag). In cases where optical identifications are available, spectroscopic redshifts can be obtained either from previous JWST measurements or through follow-up observations using JWST-NIRSPEC for fainter sources. Spectroscopic campaigns will also be initiated using 10m and 30m class telescopes such as Keck MOSFIRE, Subaru PFS, or the ELT. Additionally, grism spectroscopic campaigns conducted by Roman and Euclid will provide valuable sources of redshifts for the brightest sources with redshifts z$<$6-8.

Spectra play a crucial role in determining the masses of black holes, which is essential for understanding the distributions of Eddington rates and tracking the evolution of the M$_{BH}-\sigma$ relation. By obtaining spectroscopic information, AXIS, in conjunction with other observatories and telescopes, will contribute to advancing our knowledge of the masses of black holes in high-redshift AGNs. This, in turn, will enable detailed investigations into the accretion processes and the relationships between black hole masses and the properties of their host galaxies.

To assemble an even larger sample of redshift measurements, photometric redshifts (Photo-z) will be relied upon. Photo-z has become a well-established technique to determine the redshifts of AGNs \cite{salvato,ananna17}. The abundant amount of ancillary photometric data available from NIR surveys will enable precise measurements of Photo-z. Using this photometric information, we can obtain high-quality estimates of redshifts for almost all sources detected in the survey.

The Photo-z measurements can be effectively employed in determining the X-ray luminosity function (XLF). The sources can be distributed across the redshift range based on their associated probability density functions, allowing for a comprehensive determination of the XLF \cite[e.g.,][]{ananna17}. This approach enables us to derive statistical information about the AGN population and its evolution over cosmic time.

As a valuable by-product of this work, we will obtain measurements of the Spectral Energy Distribution (SED) for all sources. The SEDs provide crucial insights into the host properties of AGNs, including stellar mass, gas mass, extinction, and indirectly the black hole mass and Eddington rates. By analyzing the SEDs, we can gain a deeper understanding of the physical characteristics and compositions of the host galaxies, shedding light on the intricate interplay between the central supermassive black hole and its environment.

Through the utilization of photometric redshifts and the comprehensive analysis of SEDs, AXIS will significantly contribute to the study of AGN populations, their properties, and their coevolution with host galaxies. This approach will provide valuable information to advance our understanding of the growth and dynamics of supermassive black holes across cosmic time.

\subsection{Stacking Analysis of JWST detected high-z galaxies}
The strategic stacking of X-ray images at locations of known galaxies offers a remarkable increase in sensitivity. This approach provides a powerful means to investigate the average properties of a population that may otherwise be challenging to study individually. An illustrative example of the efficacy of this technique was demonstrated by \cite{vito18}, who stacked X-ray signals from HST-selected massive galaxies at redshifts of 4-5 using the 7 Ms CDFS (Chandra Deep Field South). Using this method, they achieved effective exposure times of approximately 10$^9$ s, exceeding Chandra's capabilities, and successfully detected the cumulative X-ray signal up to redshifts around 5.

Considering a short AXIS exposure of 50 ks in regions covered by JWST, assuming a magnitude limit of 28th at 7 $\mu$m, it is estimated that the expected source density would be on the order of 20000 deg$^{-2}$. If 1\% of these sources are AGN, statistically significant fluxes on the order of 5$\times$10$^{-19}$ \fcgs can be achieved. At a redshift of approximately 4, this corresponds to luminosities of around 10$^{41}$ \lx. In the deep fields, AXIS will reach even deeper flux levels, although the scaling with exposure time is not linear because of the transition from being photon-limited to being background-limited at such depths. This capability will enable AXIS to study populations of very faint sources detected by JWST, extending investigations to redshifts well beyond 10.

By employing the stacking technique and leveraging the synergistic observations with JWST, AXIS will significantly contribute to our understanding of the high-redshift universe and the properties of faint AGN populations. This combined effort will push the boundaries of knowledge, enabling investigations into the early stages of galaxy and black hole formation and evolution.

\section{Conclusions}

In this white paper, we have thoroughly explored the necessity and potential of AXIS in investigating the early accretion of supermassive black holes (SMBHs) in the universe. Specifically:
\begin{itemize}
\item AXIS will measure the X-ray Luminosity Function (XLF) of AGN up to redshifts around 10, providing crucial insights into their evolution over cosmic time with its deep and intermediate surveys in combination with ancillary information from JWST, Roman, Euclid, LSST-Rubin, and the ELT. With the XLF, AXIS will inform and constrain models of SMBH seeding, potentially helping to distinguish between scenarios where they grow from light or heavy seeds.
\item AXIS will significantly expand the current flux-area parameter space of X-ray surveys, covering a range of 0.13-2.5 deg$^2$, and reaching much fainter fluxes. This, coupled with the constant Point Spread Function (PSF) across the field of view, will result in a leap forward in survey grasp, achieving a two-order-of-magnitude improvement compared to Chandra.
\item AXIS will play a crucial role in determining the nature of high-redshift galaxies observed by JWST that exhibit characteristics typical of star-forming galaxies (SFGs), aiding in their classification as AGN in particular in the case of highly obscured and low-metallicity sources.
\item The excellent angular resolution of AXIS will facilitate clear identification of X-ray sources with NIR detections, enabling comprehensive studies of their Spectral Energy Distributions (SEDs) and spectral features.
\item With access to a wealth of ancillary data, AXIS will enable the study of the onset and evolution of galaxy-SMBH coevolution.
\item The high source surface density observed by AXIS will allow for the characterization of the coevolution of early AGN with their environments through clustering studies. It will also enable investigations into the environment of individual high-redshift quasars.
\item Leveraging the low background and exceptional angular resolution, stacking AXIS images at the positions of very faint JWST sources will facilitate the study and constraint of the AGN population, potentially extending to redshifts beyond 10 and probing low-luminosity (black hole mass) sources at lower redshifts.
\end{itemize}

This white paper, along with the entire series, emphasizes the crucial and unparalleled importance of adopting a high-angular-resolution X-ray telescope such as AXIS as a key instrument for studying the Universe in the early 2030s. AXIS, a remarkable X-ray survey machine building upon the legacy of Chandra, will surpass the achievements of its predecessor and push the boundaries of black hole exploration into new and uncharted territories.





\vspace{6pt} 



\acknowledgments{We kindly acknowledge the entire AXIS team for outstanding scientific and technical work. This work is the result of several months of discussion in the AXIS-AGN SWG. NC thanks the University of Miami for support during the proposal and this paper writing phase.}


\appendixtitles{no} 
\appendixsections{multiple} 



\externalbibliography{yes}
\bibliography{references}

\end{document}